# Super-resolution MRI through Deep Learning

Qing Lyu[1,3], Chenyu You[2,3], Hongming Shan[1], Ge Wang[1]*

1 Biomedical Imaging Center, Department of Biomedical Engineering, Rensselaer Polytechnic Institute, Troy, New York, USA, 12180
2 Department of Bioengineering and Electrical Engineering, Stanford University, Stanford, California, USA, 94305
3 Co-first author

*Abstract* **–** Magnetic resonance imaging (MRI) is extensively used for diagnosis and image-guided therapeutics. Due to hardware, physical and physiological limitations, acquisition of high-resolution MRI data takes long scan time at high system cost, and could be limited to low spatial coverage and also subject to motion artifacts. Super-resolution MRI can be achieved with deep learning, which is a promising approach and has a great potential for preclinical and clinical imaging. Compared with polynomial interpolation or sparse-coding algorithms, deep learning extracts prior knowledge from big data and produces superior MRI images from a low-resolution counterpart. In this paper, we adapt two state-of-the-art neural network models for CT denoising and deblurring, transfer them for super-resolution MRI, and demonstrate encouraging super-resolution MRI results toward two-fold resolution enhancement.

*Key Words* **–** Magnetic resonance imaging (MRI), deep learning, super-resolution.

## I. Introduction

Magnetic resonance imaging (MRI) is one of the most important diagnostic imaging modalities, and reconstructs images in multi-contrasts involving proton density, T1, T2, and magnetic susceptibility. As a result, MRI reveals anatomical, physiological, cellular and molecular information, and plays an instrumental role for both preclinical and clinical studies. As for other imaging modalities, higher resolution (HR) MRI images are always desirable. However, the MRI physics severely constraints how high image resolution can be achieved with a given scanner within a practical time frame. The MRI imaging process generally requires longer scanning time than CT scanning; for example, a clinical MRI scan may take 20 minutes or even longer for millimeter-sized image resolution [1], while a routine CT scan can be done in seconds with sub-millimeter image resolution. Higher-resolution MRI can be implemented with a stronger background magnetic field and associated pulse sequences at much higher instrumentation and operational costs. If super-resolution (SR) MRI can be computationally achieved from low-resolution (LR) MRI images, regular and low-end MRI scanners will deliver super values to hospitals, clinics, and patients.

Various algorithms for super-resolution imaging were reported in the literature. As the simplest approach, conventional interpolation methods such as nearest neighbor, bilinear, and bicubic interpolations were used to magnify a LR image into a smoothly rendered version. However, these interpolation results induce artifacts [2]. Iterative back-projection algorithm proposed by Irani et al. [3-5] has been used in MRI SR. It is simple to implement but may not give a unique solution due to the ill-posed nature of the inverse problem. Much more sophisticated sparse-coding super-resolution algorithms find a sparse representation and enable a super-resolution gain, such as through dictionary learning [6]. Due to limited resolution improvement and slow execution speed in the case of 3D MRI images, this dictionary learning method has not obtained a traction for medical imaging [7]. Furthermore, priori-information-based approaches like deterministic regularized approach [8-14], statistical regularized approach [15-18] were also used for SR MRI.

Deep learning is an emerging area of research with major implications in many application domains including MRI. Over the past several years, deep-learning-based imaging has been widely investigated, improving the image quality in many aspects including denoising, artifact reduction, system calibration, and sparse-data-based reconstruction [19-25]. For super-resolution imaging, deep learning techniques were recently developed for CT with a great success [26]. For super-resolution MRI, there were some early attempts with a 3D convolutional neural network, generative adversarial network (GAN), and densely connected network [7, 27, 28].

In this paper, we adapt two state-of-the-art neural networks – conveying path-based convolutional encoder-decoder with VGG (GAN-CPCE) [23] and GAN constrained by the identical, residual, and cycle learning ensemble (GAN-CIRCLE) [26] both of which were initially developed for CT – and develop two networks for super-resolution MRI. In our simulation study, these two networks have shown a great ability in producing HR MRI images from LR counterparts. At this stage of investigation, our data indicate a 2-fold resolution improvement.

## II. Methodology

Assuming that $I_{LR}, I_{HR} \in \mathbb{R}^{m \times n}$ are respectively LR and HR MRI images of size $m \times n$ and that their relationship can be expressed as $I_{LR} = f(I_{HR})$, where $f: I_{HR} \in \mathbb{R}^{m \times n} \to I_{LR} \in \mathbb{R}^{m \times n}$ denotes the down-sampling/blurring process that creates a LR counterpart from a HR image. The super-resolution imaging process is to implement an approximate inverse function $g \approx f^{-1}: I_{HR} \approx g(I_{LR})$.

### II.1. Training and Testing Datasets

The data used in this study was from the IXI dataset [29]. Totally, 8,955 T2-weighted MRI brain slices of 256 × 256 pixels were scanned by the Philips Medical Systems Gyroscan Intera 1.5T scanner, and used as the HR labels. When creating LR images, HR images were 2D Fourier transformed and the data in the frequency domain were kept to the central 25% region. Peripheral higher frequency data were completely cut off. In this way, the LR images were made to a 2-fold lower spatial resolution than that of the HR images. In our pilot study, 8,000 slices were selected for training, and the other 955 slices for testing.

### II.2. Neural networks

First, our previously published GAN-CPCE is an improved GAN with the Wasserstein distance in combination with the VGG network for perceptual similarity measure [23]. The structure of the generator G of the GAN-CPCE is shown in Fig. 1. It has four convolutional layers with 32 3 × 3 filters, three deconvolutional layers with 32 3 × 3 filters, and one deconvolutional layer with one 3 × 3 filter. The stride of all filters is 1. The conveying paths copy early feature maps and concatenate them into later layers for preserving details of the HR image features. A pre-trained VGG-19 network [30] gives the feature map $\phi$. The discriminator D has 6 convolutional layers with 64, 64, 128, 128, 256, and 256 3×3 filters respectively, followed by 2 fully-connected layers of sizes 1024 and 1. Each layer is equipped with the leaky ReLU. The stride is 1 for odd convolutional layers and 2 for even layers. The objective function balances the Wasserstein GAN adversarial loss $\mathcal{L}_a$ [31] and the perceptual loss $\mathcal{L}_p$ [30]:

$$\min_{\theta_G} \mathcal{L} = \mathcal{L}_a + \lambda_p \mathcal{L}_p$$
$$\min_{\theta_G} \mathcal{L}_a = \mathbb{E}_{I_{LR}}[D(G(I_{LR}))]$$
$$\min_{\theta_G} \mathcal{L}_p = \mathbb{E}_{I_{LR}, I_{HR}} \| \phi[G(I_{LR})] - \phi[G(I_{HR})] \|_2^2$$

In the experiments, the feature map $\phi$ takes the 16$^{th}$ convolutional layer in the VGG network.

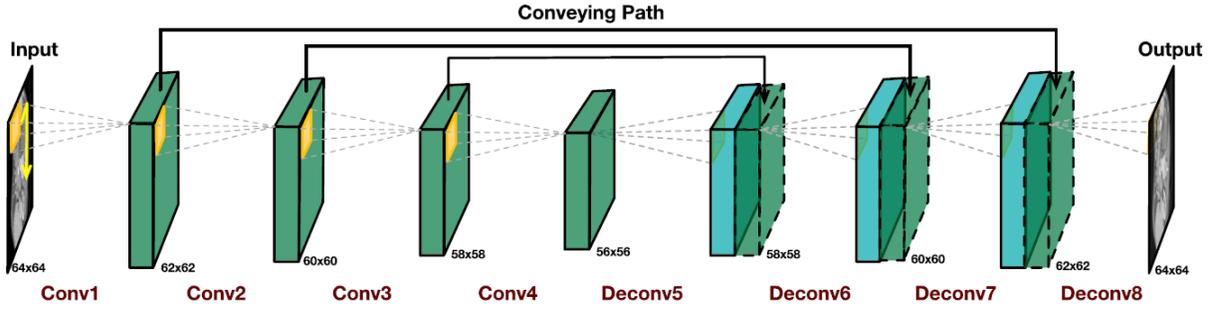

Fig. 1. The structure of the generator of GAN-CPCE from [23]. For info on the whole network, please read [23].

Second, our recent work on GAN-CIRCLE was focused on super-resolution CT. GAN-CIRCLE includes two generative networks G and F as shown in Fig. 2. Each generative network consists of a feature extraction network and a reconstruction network. In the feature extraction network, there are 12 sets of non-linear super-resolution feature blocks composed of $3 \times 3$ filters, leaky ReLU, and a dropout layer. In the reconstruction network, there are two branches in the initial part and then those two parts are integrated. These parallel CNNs utilize shallow multilayer perceptrons to perform a nonlinear estimation in the spatial domain. The structure of the discriminator is shown in Fig. 3, with 4 stages of convolution, bias, instance norm, and leaky ReLU, followed by two fully-connected layers.

The objective function of GAN-CIRCLE is composite, consisting of four parts: the adversarial loss $\mathcal{L}_{WGAN}$, the cycle consistency loss $\mathcal{L}_{CYC}$, the identity loss $\mathcal{L}_{IDT}$, and the joint sparsifying transform loss $\mathcal{L}_{JST}$ (For details, please refer to [26]):

$$\mathcal{L}_{GAN-CIRCLE} = \mathcal{L}_{WGAN}(D_Y, G) + \mathcal{L}_{WGAN}(D_x, F) + \lambda_1 \mathcal{L}_{CYC}(G,F) + \lambda_2 \mathcal{L}_{IDT}(G,F) + \lambda_3 \mathcal{L}_{JST}(G)$$

$$\min_G \max_{D_Y} \mathcal{L}_{WGAN}(D_Y, G) = -\mathbb{E}_y[D(y)] + \mathbb{E}_x[D(G(x))] + \lambda \mathbb{E}_{\tilde{y}}[(\| \nabla_{\tilde{y}} D(\tilde{y}) \|_2 - 1)^2]$$

$$\mathcal{L}_{CYC}(G,F) = \mathbb{E}_x[\| F(G(x)) - x \|_1] + \mathbb{E}_y[\| G(F(y)) - y \|_1]$$

$$\mathcal{L}_{IDT}(G,F) = \mathbb{E}_y[\| G(y) - y \|_1] + \mathbb{E}_x[\| F(x) - x \|_1]$$

$$\mathcal{L}_{JST}(G) = \tau \| G(x) \|_{TV} + (1-\tau) \| y - G(x) \|_{TV}$$

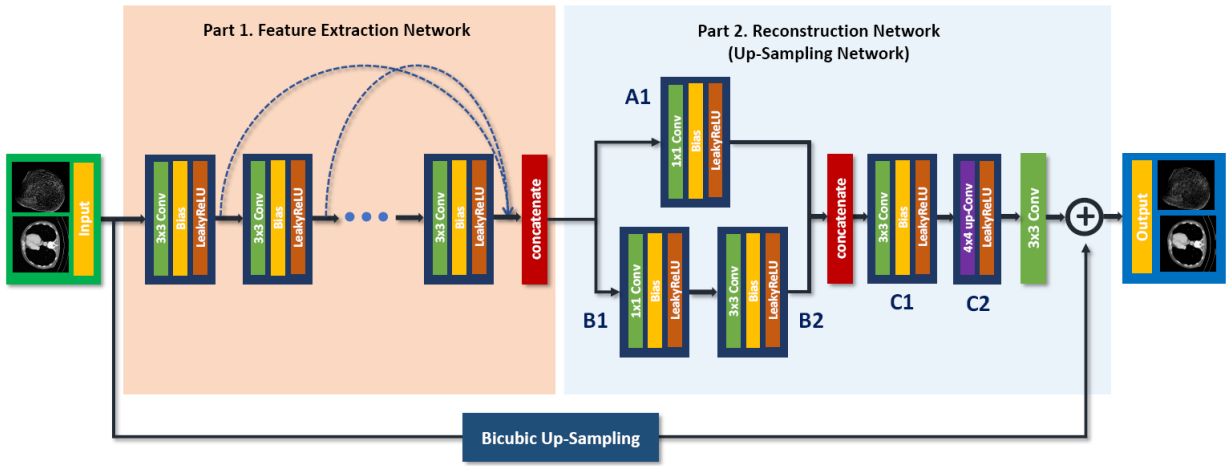

Fig. 2. The structure of the generator in the GAN-CIRCLE, copied from [26].

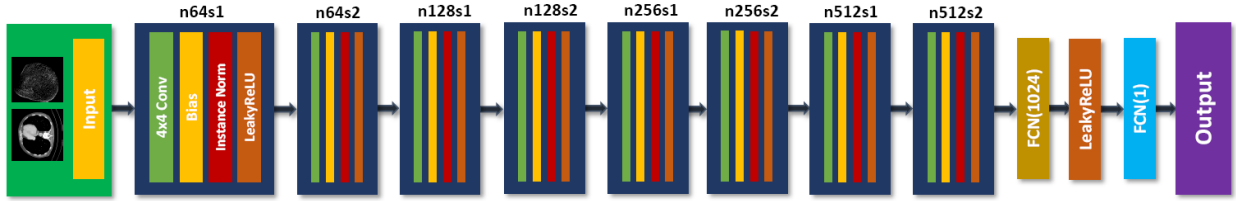

*Fig. 3. The structure of the discriminator in the GAN-CIRCLE, referred from [26].*

## II.3. Training Process

During the training process, the LR and HR images were divided into 64 × 64 patches. The training process was conducted in the supervised learning framework. For both for the two neural networks, the training process continued for 100 epochs with the learning rate of 0.0001 in the first fifty epochs and then linearly reduced to 0.000002 in the later fifty epochs. Cross-validations were also performed for each of the adapted networks.

## III. Results and Discussions

After training for 100 epochs, typical testing results from the two SR MRI imaging neural networks are in Figs. 4 and 5 respectively. Compared to the LR images, the learned SR MRI images from either of our neural networks showed encouraging improvements. By looking at the zoomed regions of interest (the red block), it can be found that the boundaries of cerebral sulcus become much clearer than those in the LR images, in an excellent agreement with the HR images.

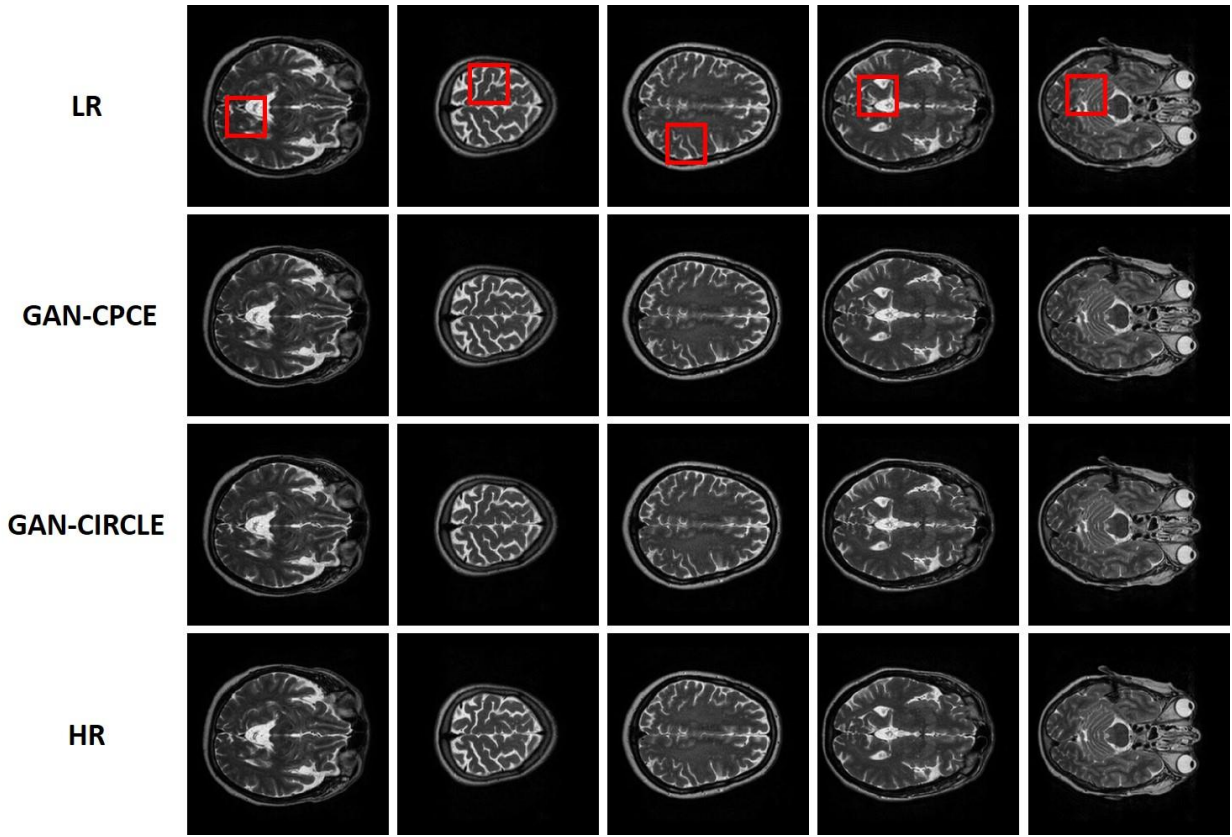

*Fig. 4. Representative SR MRI imaging results with GAN-CPCE and GAN-CIRCLE respectively.*

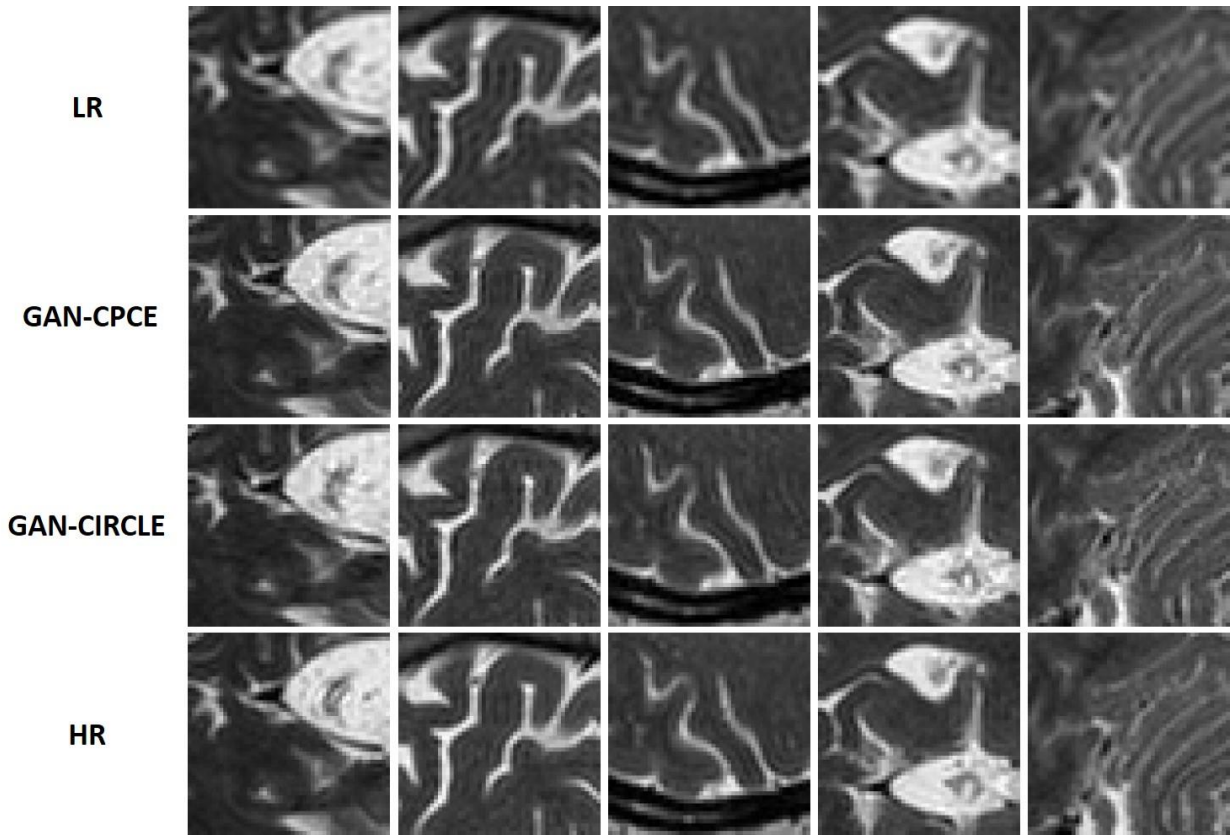

*Fig. 5. Detailed comparison in the zoomed regions of interest marked as the red boxes in Fig. 4.*

Quantitatively, the structural similarity (SSIM) and peak signal-to-noise (PSNR) were used to evaluate the image quality improvements with SR learning. Table 1 shows that the SSIM and PSNR measures of the enhanced MRI images from both neural networks. Compared GAN-CIRCLE, GAN-CPCE produced higher SSIM and PSNR values in this pilot study. All these values are signficantly better than that of the LR images.

*Table 1. SSIM and PSNR of result images reconstructed from GAN-CPCE and GAN-CIRCLE networks.*

|      | *GAN-CPCE* | *GAN-CIRCLE* |
|------|------------|--------------|
| **SSIM** | 0.9563 | 0.9465 |
| **PSNR** | 33.2031 | 32.2995 |

Further refinements are under way. Specifically, several changes will be made to improve the performance of the two neural networks for MRI deblurring. First, we will enlarge the training dataset to improve the generalization ability of the neural networks. Second, we will continue refining hyper-parameters for training. Then, we will combine the loss terms in different ways to improve the overall super-resolution performance in a task-specific fashion. Finally, we will conduct not only supervised training but also semi-supervised and unsupervised training.